\documentclass[aps,prl,twocolumn,groupedaddress]{revtex4-2}
\usepackage{aas_macros}
\usepackage{tikz}
\usepackage{amsmath}
\usepackage{bm}
\usepackage{makecell}
\DeclareMathOperator{\sech}{sech}
\usepackage{xcolor}

\begin{document}


\title{
Comment on ``Nonideal Fields Solve the Injection Problem in Relativistic Reconnection''}


\author{Fan Guo}
\affiliation{Los Alamos National Laboratory, Los Alamos, NM 87545, USA}
\email[]{guofan@lanl.gov}

\author{Xiaocan Li}
\affiliation{Dartmouth College, Hanover, NH 03755, USA}
\author{Omar French}
\affiliation{Department of Physics, 390 UCB, University of Colorado, Boulder, CO 80309, USA}

\author{William Daughton}
\affiliation{Los Alamos National Laboratory, Los Alamos, NM 87545, USA}

\author{William Matthaeus}
\affiliation{Department of Physics and Astronomy, University of Delaware, Newark, DE 19716, USA}

\author{Qile Zhang}
\affiliation{Los Alamos National Laboratory, Los Alamos, NM 87545, USA}

\author{Yi-Hsin Liu}
\affiliation{Dartmouth College, Hanover, NH 03755, USA}

\author{Patrick Kilian}
\affiliation{Space Science Institute,
 4765 Walnut St, Suite B, Boulder, CO 80301}

\author{Grant Johnson}
\affiliation{Princeton Plasma Physics Laboratory
100 Stellarator Rd, Princeton, NJ 08540}

\author{Hui Li}
\affiliation{Los Alamos National Laboratory, Los Alamos, NM 87545, USA}



\maketitle




In a recent Letter, Sironi \cite{Sironi2022} (S22) reported the correlation between particles accelerated into high energy and their crossings of regions with electric field larger than magnetic field ($E>B$ regions) in kinetic simulations of relativistic magnetic reconnection \cite{Sironi2014,Guo2014,Guo2015,Werner2016,Guo2019,Guo2020}. They claim that electric fields in $E>B$ regions (for a vanishing guide field) dominate in accelerating particles to the injection energy $\gamma_{\rm inj}\sim\sigma$ (magnetization). They suggest that the diffusion regions host particles for a sufficient time for efficient injection. 
S22 presented test-particle simulations showing that if particle energies are reset to low energies in $E>B$ regions, efficient injection is suppressed. This issue has strong implications for modeling large-scale reconnection system, and thus important to resolve.

This Comment re-examines these claims by analyzing a simulation resembling the reference case in S22 using VPIC \citep{Bowers2008}. We find nearly \textit{no} particle stayed in $E>B$ regions long enough to achieve injection by the reconnecting electric field. 
The acceleration in $E>B$ regions only contributes a small fraction to the injection energy ($\sim10\%\gamma_{\rm inj}$ on average) \citep{Guo2019}. The energization \textit{before} any $E>B$ crossings has a comparable contribution, indicating $E>B$ regions are not unique in pre-accelerating particles. A new test-particle simulation shows that zero-outing electric fields in $E>B$ regions does not strongly influence the injection. We suggest that the procedure used in S22 to exclude $E>B$ acceleration partly removes acceleration outside $E>B$ regions, leading to a false conclusion that injection by $E>B$ regions is a necessary prerequisite.

\begin{figure*}
\includegraphics[width=0.92\textwidth]{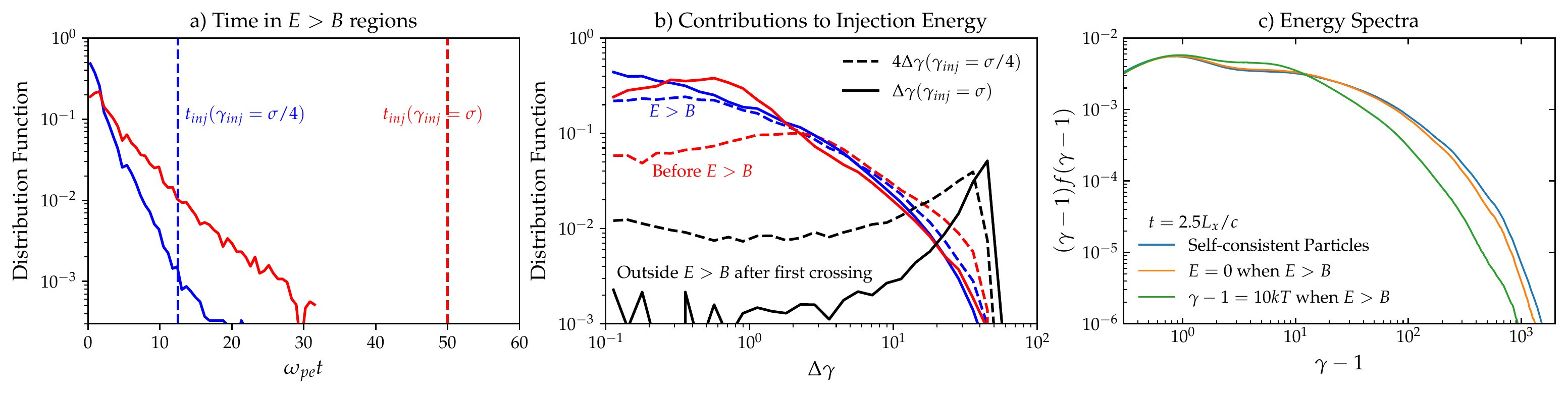}
\caption{a) Distribution function for the time duration of $E>B$ particles within $E>B$ regions. The estimated time limits for achieving injection are labeled; 
b) Distribution of energy gain during injection (for $E>B$ particles) for $E>B$ regions (blue), before $E>B$ crossing (red), and outside $E>B$ regions after the first crossing (black); c) Spectra for self-consistent particles (blue), test-particles that do not see electric fields in $E>B$ regions (red) and test-particles with a resetting energy approach (green; resembling S22). \label{fig1}}
\end{figure*}

 The initial magnetic field $\mathbf{B}=B_0\tanh(z/\lambda)\mathbf{e_x}+B_0\sech(z/\lambda)\mathbf{e_y}$. $B_0$ is the reconnecting-field magnitude and $\lambda (= 6$ skin depth $d_e$) is the half-layer-thickness. $\sigma=50$ and temperature $kT=0.36m_ec^2$. The box size is $L_x\times L_z=1600d_e\times1200d_e$ and the simulation lasts $2.5L_x/c$, with a small perturbation added to trigger reconnection. Each $d_e$ is resolved by $4$ cells with $100$ positron-electron-pairs per cell. Boundary conditions are periodic in the $x$-direction and conducting (reflecting) in the $z$-direction for fields (particles). We trace $1.28$ million particles uniformly and record the electromagnetic fields they experience at every time-step \cite{Guo2021}.
The reconnection dynamics and nonthermal energy spectra in the magnetically-dominated regime have been well documented \cite{Sironi2014,Guo2014,Guo2015,Werner2016,Guo2019,Guo2020,Sironi2016}.

During injection $\gamma\rightarrow\sigma$($\sigma/4$), $77.5\%$($51.0\%$) of the injected tracers have $E>B$ crossings (``$E>B$ particles''). S22 found a stronger correlation, since they label all particles that ever crossed $E>B$ regions during the entire simulation, rather than just during injection \cite{Sironi2022b}. 
 Clearly, there is a significant fraction of particles injected without the need to cross the $E>B$ regions \cite{French2022}. Nevertheless, it is still interesting to explore if $E>B$ regions are important for particles that crossed those regions before achieving injection.

During injection, $E>B$ particles can have multiple $E>B$ crossings. Our analysis includes all the duration that particles are in $E>B$ regions.
The time tells the limit of acceleration in the regions $\Delta \gamma_{E>B} \lesssim \int q r B_0 c dt / (m_ec^2)$, where reconnection rate $r \sim 0.1$. For $\sigma = 50$, $\omega_{\rm pe}t_{\rm inj}\gtrsim50$ is needed for $\gamma_{\rm inj}=\sigma$ ($\omega_{\rm pe}t_{\rm inj}\gtrsim12.5$ for $\gamma_{\rm inj}=\sigma/4$). Fig. \ref{fig1}a shows the time distribution of $E>B$ particles stayed in $E>B$ regions during injection. 
The mean time that particles stay in those regions is $\omega_{\rm pe}\bar{t}=4.2(1.8)$ for $\gamma_{\rm inj}=\sigma$($\sigma/4$) and nearly \textit{no} $E>B$ particles have time to reach $\gamma_{\rm inj}$. Fig. \ref{fig1}b shows the distribution of particle energy gain (during injection) in $E>B$ regions (blue), before $E>B$ crossings (red), and outside $E>B$ regions after the first $E>B$ crossing (black). Consistently, the acceleration in $E>B$ regions is too little for direct injections, with $\Delta\bar{\gamma}_{E>B}=4.8(1.6)$ for $\gamma_{\rm inj}=\sigma(\sigma/4)$. Interestingly, we find comparable acceleration before particles encounter any $E>B$, giving $\Delta\bar{\gamma}_{b,E>B}=5.6(2.5)$ for $\gamma_{\rm inj}=\sigma(\sigma/4)$. This suggests that $E>B$ acceleration is not unique in pre-accelerating particles. 
Note that this result is consistent with Fig. 3 in S22, but unfortunately overlooked in their interpretation. Fig. \ref{fig1}b also shows that most acceleration during injection occurs outside $E>B$ regions. We evolve a test-particle component in the simulation that does not “see” the electric field in $E>B$ regions (so no acceleration during each crossing), and find $84\%$ ($94\%$) for $\gamma_{\rm inj}=\sigma$($\sigma/4$) compare to self-consistent particles are still injected. There is no significant difference between energy spectra of the test-particles and self-consistent particles (Fig. \ref{fig1}c). In contrast, when particle energies are reset to an energy of $10kT$ during $E>B$ crossings (resembling S22), particle injection is suppressed. Obviously, this difference is because resetting particle energy removes the acceleration before and between $E>B$ crossings.

Our analysis demonstrated that the apparent correlation between particle injection and $E>B$ crossings do not have direct physical relation. Most acceleration for $E>B$ particles is not achieved by $E>B$ regions. We have reached the same conclusion for different $\sigma$ and domain sizes, which will be presented elsewhere.

\textit{Acknowledgment.---}
 We are thankful for discussions with Lorenzo Sironi, as well as discussions with Joel Dahlin, Jim Drake, Colby Haggerty, Dmitri Uzdensky, and Greg Werner. We acknowledge the support from Los Alamos National Laboratory through the LDRD program, DOE office of science, and NASA programs through the Astrophysical Theory Program. The work by X.L. and Y. L. is funded by the National Science Foundation grant PHY-1902867 through the NSF/DOE Partnership in Basic Plasma Science and Engineering and NASA 80NSSC21K2048.
The simulations used resources provided by the Los Alamos National Laboratory Institutional Computing Program, the National Energy Research Scientific Computing Center (NERSC) and the Texas Advanced Computing Center (TACC).

%

\end{document}